\documentclass{appolb}
\usepackage{epsfig}

\begin{document}
\title{ At the extremes of nuclear charge and spin %
\thanks{Presented by W.J. \'{S}wi\c{a}tecki at the 
XXXV Zakopane School of Physics, Zakopane, Poland, 5-13 September 
2000.
Proceedings to be published in Acta Physica Polonica.}%
}
\author{W.D. Myers and W.J. \'{S}wi\c{a}tecki
\address{Nuclear Science Division, Lawrence Berkeley 
Laboratory,\\
Berkeley, California 94720}}
\maketitle
\begin{abstract}

Using scaling rules valid in the liquid drop model of nuclei, as 
well as universal rules associated with exchanges of stability in 
families of equilibrium configurations, we constructed closed 
formulae in terms of the atomic and mass numbers Z and A and the 
angular momentum L, which represent the properties of nuclei 
rotating synchronously (with `rigid' moments of inertia), as 
calculated numerically using the Thomas-Fermi model of [5,6].  
The formulae are accurate in the range of mass numbers where the 
transition to rapidly elongating triaxial `Jacobi' shapes takes 
place.  An improved set of formulae is also provided, which takes 
account of the decreased moments of inertia at low angular 
momenta.  The formulae should be useful in guiding experimental 
searches for the Jacobi transition.  In the second part of the 
paper we discuss qualitatively some aspects of the dynamics of 
nucleus-nucleus fusion, and outline a possible way of estimating 
cross-sections for the synthesis of superheavy nuclei. 

\end{abstract}

\PACS{21.10.Dr, 21.60.Ev, 24.10.Nz}
\vspace{3ex}  

\section{Introduction}

In 1834 C.G.J. Jacobi made a startling discovery 
which led to the realisation that, at a 
certain critical angular momentum, the stable equilibrium shape of a 
gravitating mass rotating synchronously (i.e., with all mass 
elements sharing a common angular velocity) changes abruptly from 
a slightly oblate spheroid to a triaxial ellipsoid rotating about 
its shortest axis [1]. 
In 1961 the suggestion was made in [2] 
that a similar phenomenon might be expected in the case of atomic 
nuclei idealized as charged incompressible liquid drops endowed 
with a surface tension.  This was confirmed and quantified in 
1974 [3] and 1986 [4].   In 1996 the oblate-to-triaxial 
transition was demonstrated also in the more realistic
self-consistent, semi-classical nuclear Thomas-Fermi model under the 
same assumption of synchronous rotation [5].  The Thomas-Fermi 
model [6] provides a good description of shell-averaged static 
nuclear properties, but the assumption of synchronous rotation is 
known to be strongly violated at low angular momenta, where 
measured moments of inertia are considerably smaller than the 
`rigid-body' values implied by synchronous rotation [7].  In the 
first part of the present paper we provide: a) closed formulae 
that represent accurately the energies and fission barriers of 
synchronously rotating Thomas-Fermi nuclei in the range of mass 
numbers where the Jacobi transition takes place, and b) modified 
formulae that take into account the decreased moments of inertia 
at low angular momenta.

In the second part (section 5) we present a discussion of some 
aspects of the dynamics of nucleus-nucleus fusion, and we sketch 
a possible way of estimating fusion cross-sections for the 
synthesis of heavy and superheavy nuclei.

\section{Thomas-Fermi formulae}

For each of the following six nuclei: $^{74}$Se, $^{94}$Mo, 
$^{108}$Cd, $^{126}$Xe, $^{140}$Nd, $^{168}$Yb, we generated 
self-consistent stable as well as saddle-point solutions of 
rotating configurations using the Thomas-Fermi model of [5,6].  
As a rule, the angular momenta ranged between $L=0$, through 
$L=L_1$, where the Jacobi transition takes place, to $L=L_2$, 
where the barrier against fission of the Jacobi shapes vanishes.  
Using as a guide scaling rules valid in the liquid drop model 
(which is a lowest-order approximation to the Thomas-Fermi model 
[8]), as well as universal rules associated with bifurcations and 
limiting points in families of equilibrium shapes [9], we 
constructed formulae in terms of the atomic and mass numbers $Z$ 
and $A$ and the angular momentum $L$, which represent accurately 
the numerically calculated properties of the above six nuclei.  
These formulae, listed below, can then be used for neighbouring 
nuclei, thus avoiding the need for a separate Thomas-Fermi 
calculation for each additional nucleus of interest.
In the following formulae all energies are in MeV, and angular 
momenta are in units of $\hbar$.

The critical angular momentum at which the Jacobi transition 
takes place:
\begin{equation}
L_1=0.06029A^{7/6}\sqrt{40.83-\zeta}\;,
\end{equation}
where the fissility $\zeta$ is defined by
\begin{equation}
\zeta=Z^2\left/A\left[1-1.7826\left(\frac{A-
2Z}{A}\right)^2\right]\right.\;.
\end{equation}
The angular momentum at which the fission barrier vanishes:
\begin{equation}
L_2=0.09108A^{7/6}\sqrt{36.34-\zeta}\;.
\end{equation}
The energy of the oblate (Maclaurin-like) equilibrium shapes 
(with respect to the non-rotating ground state):
\begin{equation}
E_M(L)=\gamma_1 L_1 \left(0.3\lambda^2 - 0.025 
\lambda^4\right)\;,
\end{equation}
where
\begin{equation}
\gamma_1=6.2811\sqrt{(44.60-\zeta)/A}\;,
\end{equation}
and
\begin{equation}
\lambda=L/L_1\;.
\end{equation}
The energy of the Jacobi shapes (for $L_1\leq L\leq L_2$):
\begin{eqnarray}
E_J(L) &=& 0.275 \gamma_1 L_1 + 
\mbox{\small $\frac{1}{2}$} \gamma_1\left(L_2-
L_1\right)\left[\rule{0mm}{6mm}\Gamma_2(1-X)+ \right. \nonumber\\
& & \left. \mbox{\small $\frac{1}{2}$} \left(1-\Gamma_2-
\beta\right) \left(1-X^2\right)+
\mbox{\small $\frac{2}{3}$} \beta\left(1-
X^{3/2}\right) \rule{0mm}{6mm}\right]\;,
\end{eqnarray}
where
\begin{equation}
\Gamma_2=0.6118\left[1-(\zeta/38.91)^2\right]^2\left/\left[1-
(\zeta/33.49)^2\right]\right.\;,
\end{equation}
\begin{equation}
X=\left(L_2-L\right)\left/\left(L_2-L_1\right)\right.\;,
\end{equation}
\begin{equation}
\beta=0.3078\;.
\end{equation}
The energy of saddle-point shapes for $L\leq L_2$ :
\begin{eqnarray}
E_S(L) &=& 0.275\gamma_1L_1+
\mbox{\small $\frac{1}{2}$} \gamma_1\left(L_2-
L_1\right)\left[ \rule{0mm}{6mm}\Gamma_2(1-X)+ \right. \nonumber\\
& & \left. \mbox{\small $\frac{1}{2}$} \left(1-\Gamma_2-
\beta\right)\left(1-X^2\right)+
\mbox{\small $\frac{2}{3}$} \beta\left(1+X^{3/2}\right)
 \rule{0mm}{6mm}\right]\;.
\end{eqnarray}
The fission barrier for Jacobi shapes (with $L_1\leq L\leq L_2$):
\begin{equation}
B_J(L)=B_1X^{3/2}\;,
\end{equation}
where
\begin{equation}
B_1=\mbox{\small $\frac{2}{3}$} \gamma_1\beta\left(L_2-L_1\right)\;.
\end{equation}
The fission barrier for Maclaurin shapes with $L\leq L_1$:
\begin{eqnarray}
B_M(L) &=& \mbox{\small $\frac{1}{2}$} \gamma_1
\left(L_2-L_1\right)\left[\Gamma_2(1-X)+
\mbox{\small $\frac{1}{2}$} \left(1-\Gamma_2-\beta\right)\left(1-
X^2\right)+ \right. \nonumber\\
& & \left. \mbox{\small $\frac{2}{3}$} \beta\left(1+X^{3/2}\right)\right]-
\gamma_1L_1\left(0.3\lambda^2-0.025\lambda^4-0.275\right)\;.
\end{eqnarray}

Now define energy derivatives by
\begin{equation}
\gamma(L)\equiv 2\, dE(L)/dL\;,
\end{equation}
so that
\begin{equation}
\gamma_L\equiv\gamma(L-1)
\end{equation}
is an accurate approximation to a nominal quadrupole transition 
energy from the state $L$ to the state $L-2$.  Then for the 
Maclaurin shapes we have:
\begin{equation}
\gamma_M(L)=\gamma_1\left[1.2\lambda-0.2\lambda^3\right]
\end{equation}
and for the Jacobi shapes with $L_1\leq L\leq L_2$ we find
\begin{equation}
\gamma_J(L)=\gamma_1\left[\Gamma_2+(1-\Gamma_2-
\beta)X+\beta\sqrt{X}\right]\;.
\end{equation}
For $L=L_1$ we have 
$\gamma_M\left(L_1\right)=\gamma_J\left(L_1\right)=\gamma_1\;$.
For $L=L_2$ we have $\gamma_J\left(L_2\right)=\gamma_1\Gamma_2\;$.

The above equations are accurate representations of the numerical 
Thomas-Fermi solutions for mass numbers $A$ greater than about 70 
and less than about 170, or for fissilities $\zeta$ greater than 
about 15.8 and less than about 30.7.  They may also be adequate 
for $A$ less than 70, but should not be used for $A$ greater than 
about 170 (fissility greater than about 30.7). The expression for 
$B_M(L)$ may not be reliable for $L$ much below $L_1$.

Fig.1 compares the values of $L_1$ and $L_2$ with the liquid drop 
model values of [3] and with the finite range liquid drop model 
values of [4].  The curves for $L_1$ up to $A\approx170$ are 
essentially the same in all three models.  The Thomas-Fermi curve 
for $L_2$ is usually intermediate between those of the other two 
models.

Fig.2 shows the energies $E_M, E_J,E_S$ and the barrier $B$ for 
the nucleus $^{108}$Cd.

Fig.3 shows the nominal quadrupole transition energies $\gamma_L$ 
for $^{94}$Mo, $^{108}$Cd, $^{140}$Nd and $^{168}$Yb, the nuclei 
that would result after emission of four neutrons from the 
compound nuclei formed in the bombardments of $^{50}$Ti, 
$^{64}$Ni, $^{96}$Zr and $^{124}$Sn by $^{48}$Ca.  These are the 
reactions recently studied in [10].  Fig.3 implies `giant 
backbends' in the gamma ray energies $\gamma_L$ at the critical 
values given by $L=L_1+1$, where the originally increasing gamma 
ray energies suddenly begin to decrease.  This decrease is a 
hallmark of the Jacobi regime of shapes, associated with their 
rapidly increasing moments of inertia.  (Note: eqs.(15,16) imply 
that if $\gamma(L)$ has a maximum at $L_1$, then $\gamma_L$ has a 
maximum at $L_1+1$.)

\section{Modified formulae}

Measured rotational spectra correspond to energies that, for low 
angular momenta, increase considerably faster than described by 
eq.(4) or illustrated in Fig. 2.  The implied low effective 
moments of inertia are associated with nuclear pairing effects, 
and are expected to disappear at higher values of $L$ [7].  In 
particular, in the regime of the very deformed Jacobi shapes 
rotating about the shortest axis, the energy estimated using 
moments of inertia associated with synchronous rotation (`rigid' 
moments of inertia) should be relatively accurate.  We have 
accordingly modified the energy plots $E(L)$ by interpolating 
between $k$ times $E_M(L)$, Eq.\ (4), for small $L$ (where $k$ is a number 
greater than 1, which implies moments of inertia less than 1), 
and the formula for $E_J(L)$, Eq.\ (7), near $L=L_2$.  Explicitly, the 
interpolation was done as follows:
\begin{equation}
E_<=\gamma_1L_1\left[k\left(0.3\lambda^2-0.025\lambda^4\right)-
a\lambda^n\right]\;\;\;\mbox{\rm for}\; L\leq L_1\;,
\end{equation}
\begin{eqnarray}
E_> &=& 0.275\gamma_1L_1+
\mbox{\small $\frac{1}{2}$} \gamma_1\left(L_2-
L_1\right)\left[\Gamma_2(1-X)+
 \mbox{\small $\frac{1}{2}$} \left(1-\Gamma_2-
\beta\right)\left(1-X^2\right)+\right. \nonumber\\
& & \left. \mbox{\small $\frac{2}{3}$} \beta\left(1-
X^{3/2}\right)+
 bX^2\right]\;\;\;\mbox{\rm for}\;  L_1\leq L\leq L_2\;,
\end{eqnarray}
where, for a given $k$, the three quantities $n,a,b$ are 
determined by the requirement of continuity of value, slope and 
curvature at $L=L_1$.  The demand for continuity of the curvature 
is motivated by recognition of the fact that collective rotations 
about axially symmetric (Maclaurin-like) shapes do not take place 
in nuclei.  This implies that, also at low angular momenta, 
collective nuclear rotations must take place about an axis that 
is not an axis of symmetry, for example about a minor axis of a 
prolate or triaxial shape.  In that case the transition from such 
a shape to the rapidly elongating Jacobi-like shape does not involve 
a spontaneous oblate-to-triaxial symmetry breaking
, and  would be 
smooth rather than abrupt.  The associated gamma ray energies 
would now be expected to change gradually from increasing to 
decreasing functions of $L$, which implies continuity of the 
second derivative of $E(L)$.  (We shall continue to refer to the 
regime of decreasing gamma ray energies as the Jacobi regime.)

The abovementioned requirements of continuity lead to the 
following formulae for $n,a,b$: 
\begin{equation}
n=\frac{-B+\sqrt{B^2+4AC}}{2A}\;,
\end{equation}
\begin{equation}
a=\frac{A}{4+2n\kappa}\;,
\end{equation}
\begin{equation}
b=\mbox{\small $\frac{1}{2}$}(1-k+2na)\;,
\end{equation}
where
\begin{equation}
A=(1.1+\kappa)(k-1)\;,
\end{equation}
\begin{equation}
B=-A+\frac{A}{\kappa}-k(1+0.6\kappa)+\Gamma_2+\frac{\beta}{2}\;,
\end{equation}
\begin{equation}
C=\frac{2}{\kappa}\left(\frac{A}{\kappa}-A-B \right)\;,
\end{equation}
where
\begin{equation}
\kappa=\frac{L_2}{L_1}-1\;.
\end{equation}
Fig.4 illustrates the modified energies and fission barriers in 
the case of  $^{108}$Cd.  The value of $k$ was taken to be 1.5 
(see below).

The formulae for the energy derivative functions are now as 
follows:
\begin{equation}
\gamma_<(L)=\gamma_1\left[k\left(1.2\lambda-0.2\lambda^3\right)-
2na\lambda^{n-1}\right]\;\;\;\mbox{\rm for}\; L\leq L_1
\end{equation}
\begin{equation}
\gamma_>(L)=\gamma_1\left[\Gamma_2+\left(1-\Gamma_2-
\beta\right)X+\beta\sqrt{X}-2bX\right]\;\;\;\mbox{\rm for}\; 
L_1\leq L\leq L_2\;. 
\end{equation}
The Jacobi regime of decreasing values of $\gamma (L)$ begins now 
at the giant back-bend angular momentum $L_m$ (always less than 
$L_1$) where $\gamma_<(L)$ has its maximum.  It is given by the 
solution of
\begin{equation}
k\left(0.6-0.3\lambda_m^2\right)-n(n-1)a\lambda_m^{n-2}=0\;,
\end{equation}
where $\lambda_m=L_m/L_1$. 
(The maximum in $\gamma_L$ is then at $L_m+1$ --- see above.)

Varying $k$ results in a one-parameter family of interpolation 
functions for $\gamma_L$, illustrated for $^{94}$Mo in Fig.5.  
The choice $k$=1.5 leads to $\gamma_L$ plots shown in Fig.6.  
This choice turned out to give a rough correspondence with the 
preliminary results of the measurements referred to earlier [10].  
The original, unmodified curves in Fig.3 bear little resemblance 
to the data.

\section{Relation to microscopic calculations}

In a recent preprint entitled ``Very extended nuclear shapes near 
A=100'' R. R. Chasman describes a `cranked-Strutinsky' study of 37 
nuclei between $^{100}$Zr and $^{122}$Xe, at angular momenta $L=60$ 
and $L=70$ [11].  Many of these nuclei are found to have strongly 
deformed prolate or triaxial shapes, and to have fission barriers 
in the range from about 4 MeV to about 17 MeV.  The circles in 
Fig.4 show these barriers in the case of $^{108}$Cd.  For a 
sample of 17 out of the 37 cases studied by Chasman the 
deviations between the microscopic cranked-Strutinsky and the 
modified Thomas-Fermi barriers were $-0.67\pm 1.18$ MeV at $L=60$, and 
$0.41 \pm 1.53$ MeV at $L=70$.  
(The deviation for all 34 values at 
both $L=60$ and $L=70$ was $-0.13 \pm 1.45$ MeV.)  It is interesting to 
note that, since the Thomas-Fermi energies are smooth functions 
of $A$, $Z$ and $L$, one concludes that the microscopic energies are 
also smooth to within about 1.5 MeV on the average (or else that 
the shell corrections for the equilibrium shape and for the 
saddle-point shape are approximately the same).  Also, considering the 
very different inputs in the two types of calculation, it is 
remarkable that, on the average, 
the absolute values of the fission barriers agree to within
a fraction of an MeV.  On the whole, one is led to the 
conclusion that Chasman's ``very extended shapes'' and the 
Thomas-Fermi Jacobi configuarations are the microscopic and macroscopic 
descriptions of the same underlying physics that goes back to 
1834, namely: ``Sufficiently rapidly rotating fluids prefer 
elongated prolate shapes''.

\section{Fusion dynamics}

It is an elementary everyday observation that if two fluid drops 
are brought into contact, there takes place a sudden growth of 
the neck --- a snap --- characterized by a time scale much shorter 
than those typical of other collective degrees of freedom 
of the system,  such as its overall length.  The driving force for 
this snap is the great saving of surface energy achieved with 
only a minor rearrangement of the fluid's mass elements in the 
vicinity of the neck.  Thus, insofar as nuclei can be regarded as 
fluids (see below for exceptions) the dinuclear configuration of 
touching fragments is expected to be transformed rapidly into a 
mononuclear shape with about the original overall length (which 
we shall refer to as the snap length).  With reference to the 
potential energy landscape underlying the fusion process, the 
system, originally in the fusion valley, is injected into the 
vicinity of the fission valley at a point along this valley 
specified approximately by the snap length.  

Once in the fission valley, the system may find itself either 
inside or outside the saddle-point barrier guarding the compound 
nucleus against disintegration by fission.  For lighter reacting 
systems the former is the case and, after contact, fusion takes 
place automatically.  But with increasing sizes of the reacting 
nuclei the saddle-point length shrinks rapidly, so that, beyond a 
certain critical point, the situation is reversed: after the snap 
the system is outside the saddle.  The heavier the fusing 
partners the farther away from the saddle will the system find 
itself, and the greater will be the energy difference $\Delta$ 
between the saddle-point energy and the system's potential energy  
after injection into the fission valley.  This is the physics of 
the entrance channel hindrance to fusion discussed in [12].  This 
hindrance may well be the principal reason for the rapid decrease 
of measured cross-sections for the formation of very heavy 
elements.  

A second factor, {\em which acts in the opposite direction}, is 
present in the case of reactions at bombarding energies designed 
to leave the compound nucleus with a given excitation energy, for 
example the 13 MeV in the case of the reactions illustrated in 
Fig.7.  As can be seen from this figure, the typical Coulomb 
barrier in the entrance channel would prevent the relatively 
lighter projectiles up to $^{70}$Zn from even achieving contact 
between the half-density nuclear surfaces.  The implied 
hindrance, as represented by the size of the Coulomb barrier that 
protrudes above the level of the bombarding energy, is most 
pronounced for the lightest projectiles, decreasing with 
projectile size, and eventually disappearing altogether for the 
reaction $^{86}$Kr = $^{208}$Pb.  This lowering of the `Coulomb 
shield' for superheavy reactions [13,14] is an elementary 
consequence of the energetics of nuclear deformations.  Thus, the 
energy needed to deform a compound nucleus into the Coulomb 
barrier configuration of two touching fragments is resisted by the 
surface energy and favoured by the electrostatic energy. Hence, 
for a sufficiently large charge on the system, the Coulomb 
barrier will eventually sink below 
the level of the ground-state energy (or 
this energy augmented by some constant, like the 13 MeV in the 
examples above).  This is illustrated in Fig.8.

Coming back to Fig.7, the hindrance against achieving contact 
would be 100\% up to about $^{70}$Zn, and zero afterwards, if a 
classical, one-dimensional calculation were used.  In a more 
realistic treatment, the hindrance would decrease gradually, and 
a quantitative description of such `sub-barrier' fusion 
probabilities has been available for some time in terms of the 
notion of barrier height distributions [15].  

Working together with K. Siwek-Wilczy\'{n}ska and J. 
Wilczy\'{n}ski, we have been led by the above considerations to the 
following three-stage picture of the fusion process of heavy 
nuclear systems:

Stage 1: Overcoming the Coulomb barrier in order to achieve 
contact. Using existing theories of sub-barrier fusion, 
the relevant probabilities can be 
estimated.  After contact, a snap from the fusion valley into the fission 
valley follows.  The associated drop in the potential energy is 
assumed to heat up the system to a temperature T.

Stage 2: Overcoming the energy barrier $\Delta$ necessary to 
reach the compound nucleus from the fission valley after the 
snap.  We assume that this is achieved by a thermal fluctuation 
with a probability approximated by exp($-\Delta$/T).

Stage 3: Surviving the competition between fission and neutron 
emission.  Standard formulae for the relevant probabilities are 
again available.  

With the above factors depending in different ways on the 
reaction parameters, there is no particular reason to expect the 
plot of the logarithms of the formation cross-sections in Fig.7 
to continue as a linear function of the atomic number beyond 
$Z=112$.  In this 
connection it is interesting to note that the empirical data in 
Fig.7 up to $Z=112$, rather than being fitted by a straight 
line, may equally well be represented by a cubic that is made to 
pass through the point for $^{86}$Kr + $^{208}$Pb, as shown in 
Fig.9.  
 
A word about the assumption of a snap at contact, characteristic 
of fluids.  Nuclei often do exhibit fluid properties, but 
exceptions occur when sufficiently strong magic or doubly magic 
shell effects may endow a nucleus with properties of an elastic solid  
[16].  In that case the snap may not occur at contact, but only 
after a more intimate interpenetration of the partners, 
sufficient to destroy their shell effects.  Such a delay in the 
injection into the fission valley may be advantageous, since the 
resulting mononucleus will be more compact, and thus closer to 
the saddle-point configuration.  (See the discussion in [13,14].)

We hope to develop the above qualitative considerations into a 
semi-empirical method of estimating cross-sections for the
synthesis of heavy elements.

\vspace{3ex}

We would like to thank D.Ward, R.Diamond, F.Stephens, and P.Fallon  
for sharing with us information on the experimental searches 
for the Jacobi transition, and  J.Wilczy\'{n}ski and 
K. Siwek-Wilczy\'{n}ska for discussions of formation cross-sections for 
heavy nuclei. This work was supported in part by the Director, 
Office of Energy Research, Office of High Energy and Nuclear 
Physics, and by the Office of Basic Energy Sciences, Division of 
Nuclear Sciences, of the U.S. Department of Energy under
Contract No. DE-AC03-76SF00098, and by the Polish-American Maria 
Sk\l odowska-Curie \mbox{Fund No. PAA/NSF-96-253.}             

%


\begin{figure}
\begin{center}
  \begin{minipage}[t]{12.5cm}
        \epsfxsize 12.5cm \epsfbox{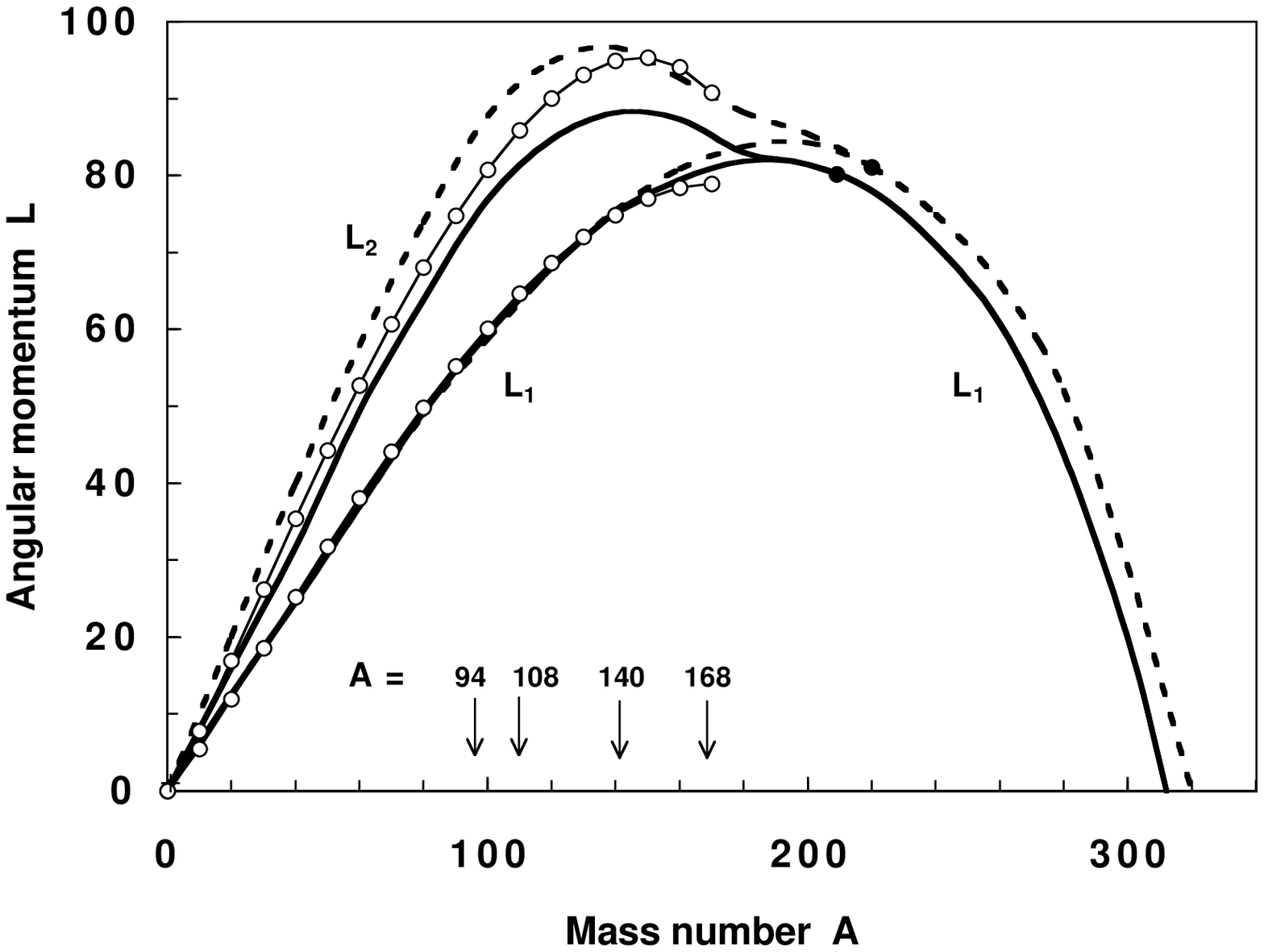}
        \hfill
    \caption{ For angular momenta below the curves labeled $L_1$ 
the equilibrium shape is an oblate configuration rotating about 
its axis of symmetry.  The three models in question are 
identified by the dashed, solid and circled curves, representing 
the liquid drop model of [3], the finite range liquid drop model 
of [4] and the present Thomas-Fermi model, respectively.  
Between $L_1$ and 
$L_2$ the equilibrium shapes are triaxial Jacobi configurations 
rotating about the shortest axis.  The curves $L_1$ and $L_2$ 
come together at the solid circles, beyond which mass numbers 
Jacobi shapes do not exist.  Disintegration takes place for 
angular momenta exceeding $L_1$ in the upper range of mass 
numbers, or $L_2$ below the solid circles.  
The curves refer to nuclei on the valley of beta stability.  The 
four arrows identify approximately the mass numbers of nuclei 
studied in [10]. 
}
  \end{minipage}
\end{center}
\end{figure}

\begin{figure}
\begin{center}
  \begin{minipage}[t]{12.5cm}
        \epsfxsize 12.5cm \epsfbox{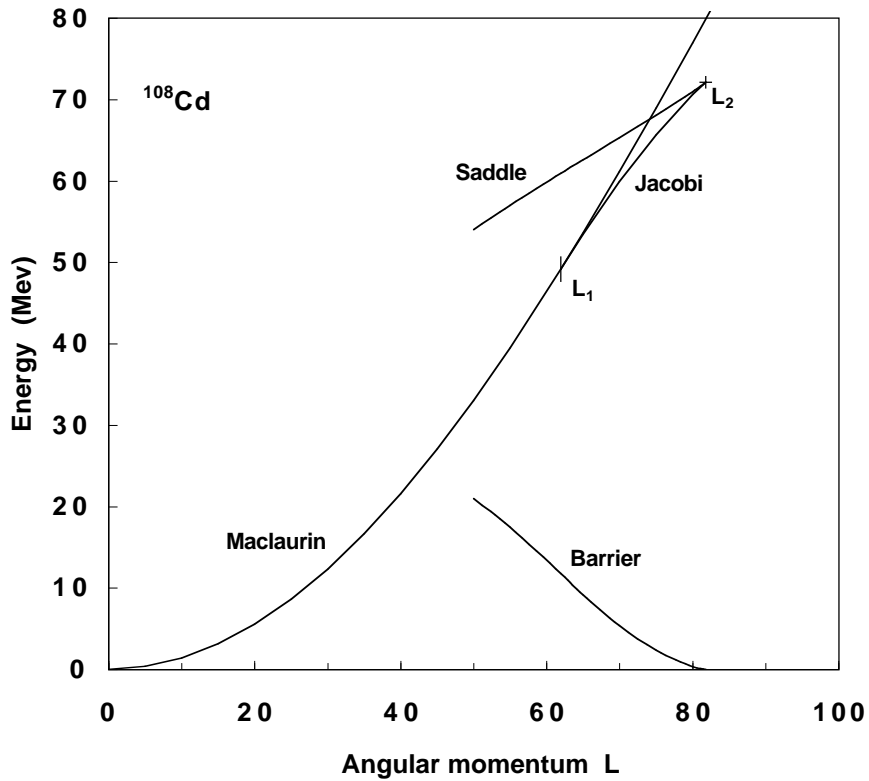}
        \hfill
    \caption{ The energies of the Maclaurin-like oblate shapes, 
the Jacobi-like triaxial shapes and the (triaxial) saddle-point 
shapes are shown in their dependence on angular momentum for 
$^{108}$Cd.  The Jacobi shapes first appear at $L_1$ and exist up 
to $L_2$.  The fission barrier $B$ is the energy difference 
between the saddle energy and either the Jacobi energy for 
$L\geq L_1$ or the Maclaurin energy for $L\leq L_1$.  It vanishes 
at $L_2$.
}
  \end{minipage}
\end{center}
\end{figure}

\begin{figure}
\begin{center}
  \begin{minipage}[t]{12.5cm}
        \epsfxsize 12.5cm \epsfbox{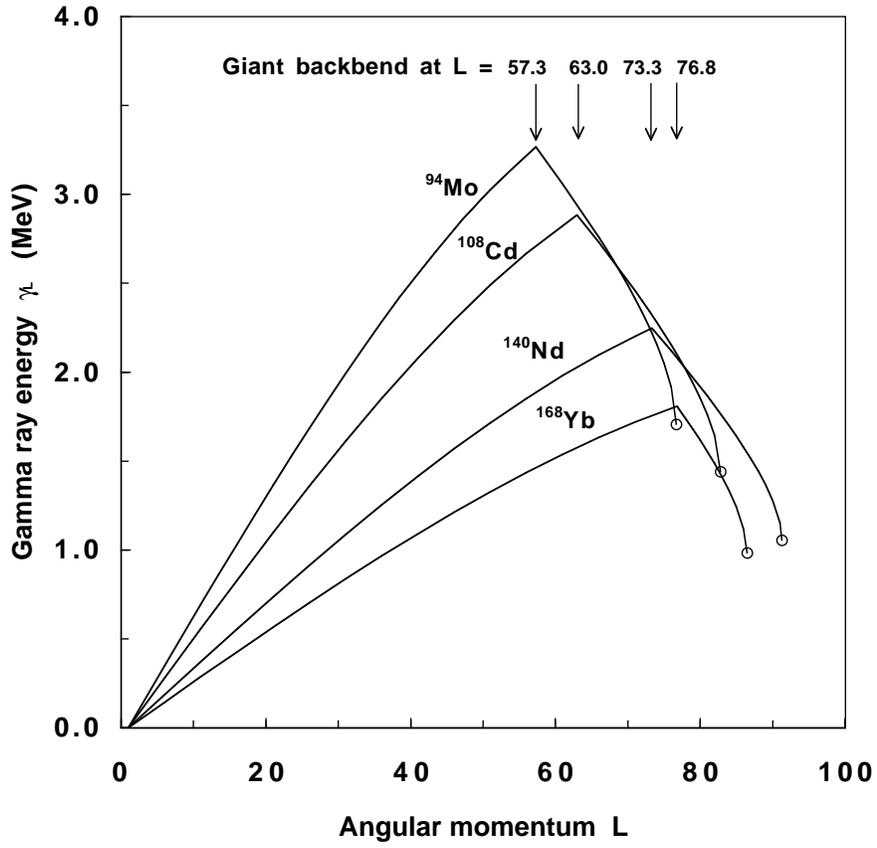}
        \hfill
    \caption{ The nominal quadrupole gamma ray energies 
$\gamma_L$ are shown in their dependence on $L$ for four nuclei.  
These curves represent the unmodified Thomas-Fermi model, with 
sharp giant backbends at the angular momenta indicated.  The 
Jacobi shapes exist beyond the backbend and terminate at the 
circled points.
}
  \end{minipage}
\end{center}
\end{figure}

\begin{figure}
\begin{center}
  \begin{minipage}[t]{12.5cm}
        \epsfxsize 12.5cm \epsfbox{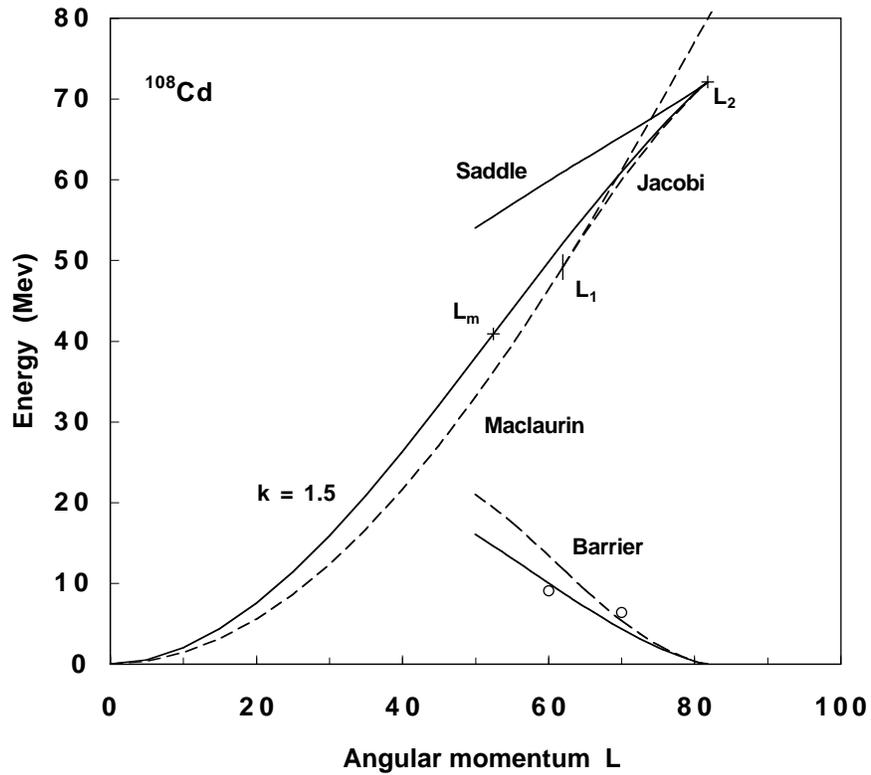}
        \hfill
    \caption{ The dashed curves repeat the plot from Fig.2, and 
the solid curves show the modification resulting from  taking 
account of the reduction of the moment of inertia at low $L$ by a 
factor of 1.5.  The Jacobi regime of decreasing
gamma ray energies begins now at the angular
momentum $L_m$.  The two circled points refer to fission barriers 
obtained with the cranked-Strutinsky method in [11].
}
  \end{minipage}
\end{center}
\end{figure}

\begin{figure}
\begin{center}
  \begin{minipage}[t]{12.5cm}
        \epsfxsize 12.5cm \epsfbox{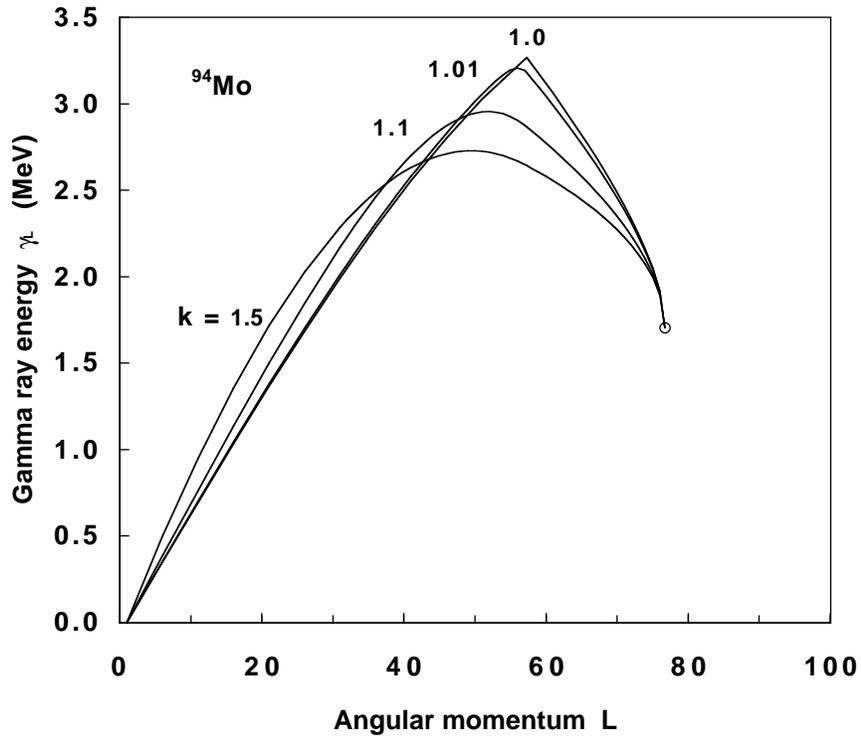}
        \hfill
    \caption{ The nominal quadrupole gamma ray energies for 
$^{94}$Mo, calculated using a modification of the Thomas-Fermi 
results, the modification consisting of assuming the low-$L$ 
moments of inertia to be reduced by 1.01, 1.1 and 1.5, 
respectively.  The curve labeled 1.0 is the unmodified Thomas-
Fermi result.  
}
  \end{minipage}
\end{center}
\end{figure}

\begin{figure}
\begin{center}
  \begin{minipage}[t]{12.5cm}
        \epsfxsize 12.5cm \epsfbox{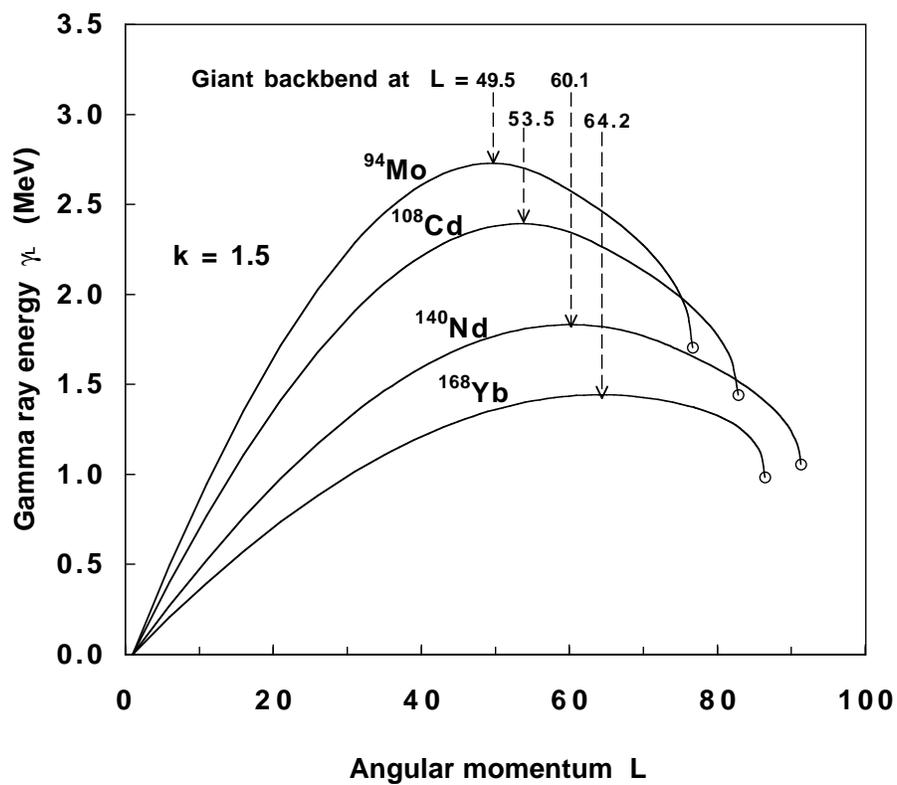}
        \hfill
    \caption{ This is like Fig.3, but after the modification 
consisting of reducing the low-$L$ moments of inertia by a factor 
1.5.  The giant backbends, marking the beginnings of the Jacobi 
regimes, are indicated by the arrows.
}
  \end{minipage}
\end{center}
\end{figure}

\begin{figure}
\begin{center}
  \begin{minipage}[t]{12.5cm}
        \epsfxsize 10cm \epsfbox{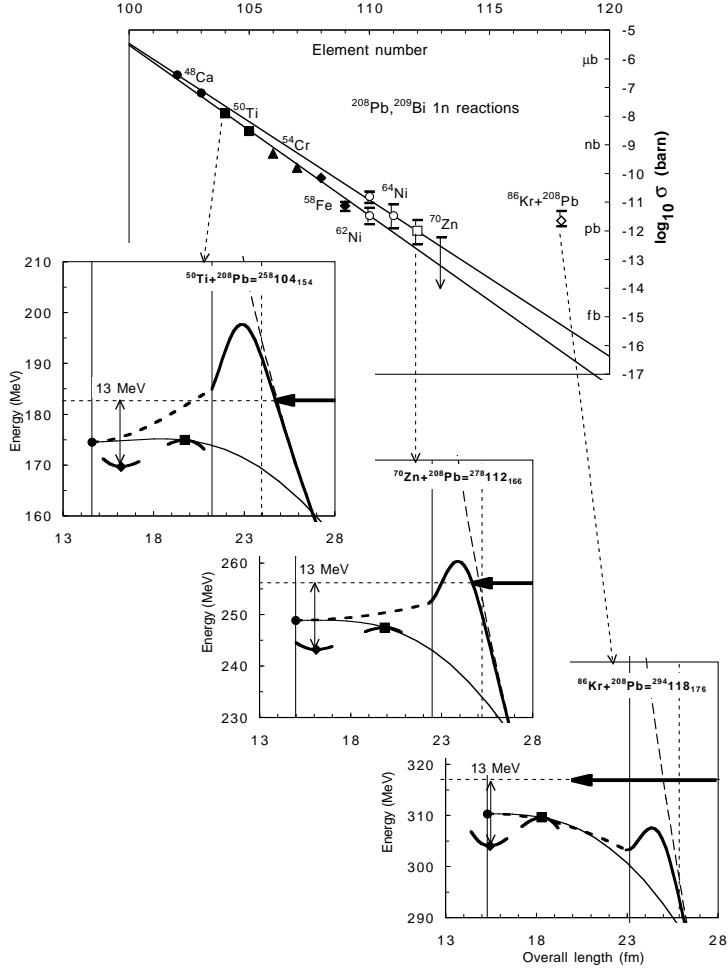}
        \hfill
    \caption{ The upper part refers to cross sections for 
synthesizing heavy elements from Z=102 to 118 in bombardments of 
$^{208}$Pb and $^{209}$Bi with projectiles from  $^{48}$Ca to 
$^{86}$Kr.  The lower part gives three examples of 
(center-of-mass) potential energy plots along the fusion valley (thick solid 
and dashed curves) and fission valley (thin curves).  The plots 
are against the overall, tip-to-tip extension of the fusing or 
fissioning configurations.  The ground states are indicated by 
diamonds, the saddle-points by squares. The solid vertical line 
corresponds to contact between the half-density radii, the dashed 
vertical line to contact of the density tails, defined by the 
classical turning points of the fastest particles in the 
approaching nuclei.  The horizontal arrow defines the bombarding 
energy, designed to leave the compound nucleus with 13 MeV of 
excitation energy.
}
  \end{minipage}
\end{center}
\end{figure}

\begin{figure}
\begin{center}
  \begin{minipage}[t]{12.5cm}
        \epsfxsize 12.5cm \epsfbox{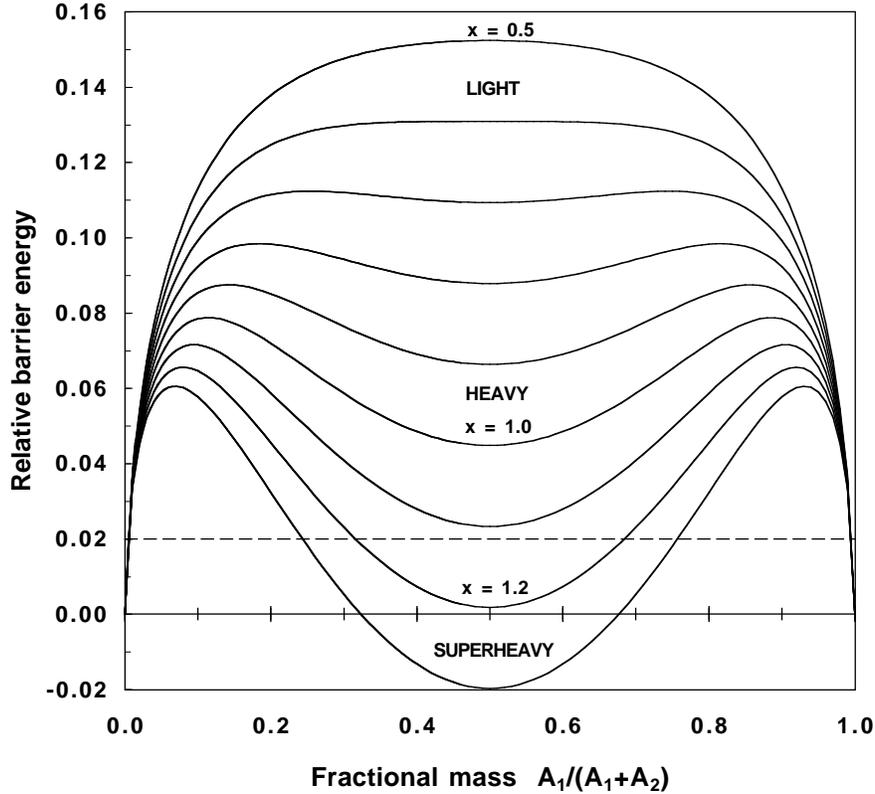}
        \hfill
    \caption{ This is the energy of tangent spheres (representing 
the Coulomb barrier) with respect to the energy of the 
single-sphere configuration (representing the compound nucleus), in 
units of that sphere's surface energy.  The plots are against the 
asymmetry of the reaction.  The label $x$ is the 
standard fissility parameter, defined as the ratio of the 
electrostatic energy of the compound sphere to twice its surface 
energy.  For $x$ slightly in excess of 1.2 the Coulomb barrier 
(in this schematic liquid drop model) sinks below the energy of 
the compound nucleus.  (This `unshielding' would occur earlier 
with respect to a somewhat higher bombarding energy that would 
allow for the
emission of one neutron.  This is indicated schematically 
by the dashed line.)  
In either case the unshielding is characteristic only of 
extremely heavy (superheavy) systems and, as a rule, would not be 
expected to have been present in most heavy nucleus-nucleus 
reactions studied so far.  (Quantitative aspects of this figure 
become modified when a more realistic macroscopic model is used, 
and when shell effects are taken into account.)  
}
  \end{minipage}
\end{center}
\end{figure}

\begin{figure}
\begin{center}
  \begin{minipage}[t]{12.5cm}
        \epsfxsize 12.5cm \epsfbox{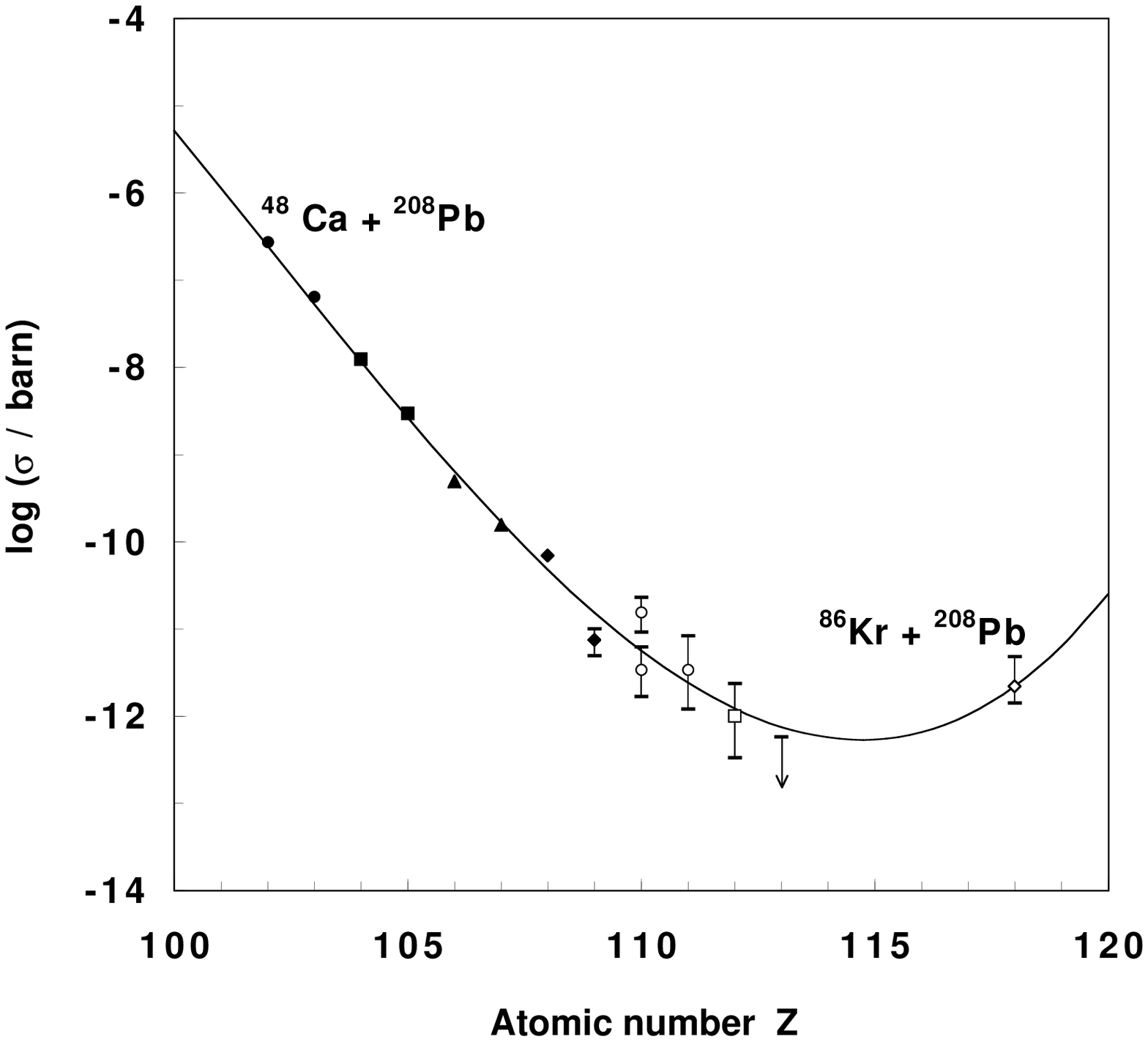}
        \hfill
    \caption{ The data points are the same as in the upper part 
of Fig.7.  The curve is the cubic given by: $\log (\sigma /
\mbox{\rm barn}) = -6.61 - 
0.6681(Z-102) +) 0.001377(Z-102)^3$.      
}
  \end{minipage}
\end{center}
\end{figure}


\begin{thebibliography}{99}

\bibitem{1}
See S.Chandrasekhar, {\em Ellipsoidal Figures of Equilibrium}, 
Yale University Press, New Haven and London, 1969.

\bibitem{2}
R.Beringer and W.J.Knox, Phys.Rev. {\bf 121}, 1195(1961).

\bibitem{3}
S.Cohen, F.Plasil and W.J.Swiatecki, Ann.Phys.(N.Y.) {\bf 82}, 
557(1974).
\bibitem{4}
A.J.Sierk, Phys.Rev.{\bf C33}, 2039(1986).

\bibitem{5}
W.D.Myers and W.J.Swiatecki, Acta Physica Polonica {\bf 27A}, 
99(1996); Nucl.Phys. {\bf A612}, 249(1997).

\bibitem{6}
W.D.Myers and W.J.Swiatecki, Nucl.Phys. {\bf A601}, 141(1996).

\bibitem{7}
A.Bohr and B.Mottelson, {\em Nuclear Strycture}, Benjamin, N.Y. 
1975; Table of Isotopes, 8th edition, ed. R.B.Firestone and 
V.S.Shirley, Wiley, N.Y. 1996; W.D.Myers and W.J.Swiatecki, 
Nucl.Phys. {\bf A641}, 203(1998), especially Figs.4-9.

\bibitem{8}
W.D.Myers and W.J.Swiatecki, Ann.Phys. (N.Y.) {\bf 55}, 
395(1969); {\bf 84}, 186(1974); W.J.Swiatecki, Nucl.Phys. {\bf 
A574}, 233c(1994).

\bibitem{9}
A general discussion of the `exchange of stability' between 
families of equilibrium shapes goes back to H.Poincar\'{e}, as 
described in P.Appell, {\em Trait\'{e} de M\'{e}canique 
Rationelle}, Gauthier-Villars, Paris 1932\@.  See also R.Thom, {\em 
Parabole et Catastrophe}, Flammarion, 1983\@.  In the case of two 
families that cross in a `pitchfork' pattern, the energy 
difference between them grows as the second power of the distance 
from the crossing; in the case of a linear crossing, it is the 
third power; in the case of a `limiting point' it is the 
three-halves power.  For some relevant examples see W.J.Swiatecki, 
Phys.Rev. {\bf 101}, 651(1956).  

\bibitem{10}
D.Ward et al., in preparation.

\bibitem{11}
R.R.Chasman, ``Very Extended Nuclear Shapes Near A=100'', Argonne 
National Laboratory Physics Division preprint PHY-9018-Th-98.
\bibitem{12}
J.B.Blocki, H.Feldmeier and W.J.Swiatecki, Nucl.Phys. {\bf 
A459}, 145(1986) and references therein.
\bibitem{13}
W.D.Myers and W.J.Swiatecki, Acta Physica Polonica {\bf B31}, 
1471(2000).
\bibitem{14}
W.D.Myers and W.J.Swiatecki, Phys.Rev. {\bf C62}, 044610(2000).
\bibitem{15}
N.Rowley, G.R.Satchler and P.H.Stelson, Phys.Lett. {\bf B254}, 
25(1991) and, for example,
J.D.Bierman et al., Phys.Rev. {\bf C54}, 3068(1996).
\bibitem{16}
J.Blocki, J.Skalski and W.J.Swiatecki, Nucl.Phys. {\bf A594}, 
137(1995); {\bf A618},1(1997) and references therein; W.J.Swiatecki,
Nucl.Phys.{\bf A488}, 375c(1988). 

\end{thebibliography}
\end{document}